\documentclass[aps,floatfix,twocolumn,amsmath]{revtex4}
\usepackage{graphicx}
\newcommand{\be}{\begin{equation}}
\newcommand{\ee}{\end{equation}}
\newcommand{\bea}{\begin{eqnarray}}
\newcommand{\eea}{\end{eqnarray}}
\newcommand{\no}{\noindent}
\begin{document}

% Use the \preprint command to place your local institutional report
% number in the upper righthand corner of the title page in preprint mode.
% Multiple \preprint commands are allowed.
% Use the 'preprintnumbers' class option to override journal defaults
% to display numbers if necessary
%\preprint{}

%Title of paper
\title{T and S dualities \\and \\The cosmological evolution of the dilaton and the scale factors}

% repeat the \author .. \affiliation  etc. as needed
% \email, \thanks, \homepage, \altaffiliation all apply to the current
% author. Explanatory text should go in the []'s, actual e-mail
% address or url should go in the {}'s for \email and \homepage.
% Please use the appropriate macro foreach each type of information

% \affiliation command applies to all authors since the last
% \affiliation command. The \affiliation command should follow the
% other information
% \affiliation can be followed by \email, \homepage, \thanks as well.
\author{Tongu\c{c} Rador}
\email[]{tonguc.rador@boun.edu.tr}

%\homepage[]{Your web page}
%\thanks{}
%\altaffiliation{}
\affiliation{Bo\~{g}azi\c{c}i University Department of Physics \\ 34342 Bebek, \.{I}stanbul, Turkey}

%Collaboration name if desired (requires use of superscriptaddress
%option in \documentclass). \noaffiliation is required (may also be
%used with the \author command).
%\collaboration can be followed by \email, \homepage, \thanks as well.
%\collaboration{}
%\noaffiliation

\date{\today}

\begin{abstract}
% insert abstract here
Cosmologically stabilizing radion along with the dilaton is one of the major concerns of low energy string theory. One can hope that T and S dualities can provide a plausible answer. In this work we study the impact of S and T duality invariances on dilaton gravity. We have shown various instances where physically interesting models arise as a result of imposing the mentioned invariances. In particular S duality has a very privileged effect in that the dilaton equations partially decouple from the evolution of the scale factors. This makes it easy to understand the general rules for the stabilization of the dilaton. We also show that certain T duality invariant actions become S duality invariance compatible. That is they mimic S duality when extra dimensions stabilize.
\end{abstract}

% insert suggested PACS numbers in braces on next line
%\pacs{02.50.-r, 05.40-a}
% insert suggested keywords - APS authors don't need to do this
%\keywords{}

%\maketitle must follow title, authors, abstract, \pacs, and \keywords
\maketitle

\section{Invitation}

Extra dimensions and the dilaton are an integral part of string theory. But in a cosmological setting one has to have them stabilized in order to recover Einstein gravity at low energies. Stabilization of extra dimensions and the dilaton has been the focus of extensive study in the literature \cite{a0}-\cite{d4}. But there is, as yet, no general concensus on the mechanism for their stabilization nor whether a single mechanism should be responsible for the stabilization of both. T and S dualities are important symmetries of string theory. In this paper we study their effect on a generalized dilaton gravity action.

Stabilizing the radion in pure Einstein gravity is not difficult. But Einstein theory is not the proper low energy limit of string theory. In reality one has to deal with the dilaton and this problem proves to be rather non-trivial. Recently an explicit model has
been presented \cite{newtr} to achieve this in a way that fits in string theory. The observation there was that since T duality plays a crucial role
in the stabilization of the radion, S duality might as well provide an understanding of the stabilization of the dilaton. In the mentioned paper it was shown that this could be the case.

Here we would like to use the same point of view but generalize the scope of the approach. We will be studying the impact of T and S dualities in a generalized dilaton theory.  

To contrast our main point clearer about the evolution of the dilaton and its stabilization we shortly review a recent
model \cite{tr3}-\cite{tr5} of the cosmology of D-branes and the dilaton. The reason we are doing this is because {\em dilaton stabilization} and that {\em a low energy theory is compatible with a constant dilaton} are rather different things. We will see that such models are a viable possibility. The action of the model in the mentined works is

\be\label{eq1}
S=\int\;dx^{d}\sqrt{-g}\;e^{-2\phi}\left[R+4(\nabla\phi)^{2}+e^{a\phi}\mathcal{L}\right]\;.
\ee

\noindent We take our metric to be (assuming all directions are flat) 

\be
ds^{2}=-dt^{2}+e^{2B(t)}\sum_{i=1}^{m}dx_{i}^{2}+e^{2C(t)}\sum_{a=1}^{p}dy_{a}^{2}\;,
\ee

\noindent with $m+p=9$ and $d=1+m+p=10$. For the D-brane Lagrangian we simply take the hydrodynamical fluid form 

\be
\mathcal{L}_{m}=-2\sum_{i}\rho_{i}\;,
\ee

\noindent where the energy densities are given as follows

\be
\rho_{i}=\rho_{i}^{o}\exp\left[-(1+\omega_{i})mB-(1+\nu_{i})pC)\right]\;\;.
\ee

The equations of motion following from this Lagrangian will be,

\begin{subequations}
\bea
\ddot{B}+k\dot{B}&=&e^{a\phi}\left[T_{\hat{b}\hat{b}}-\tau\rho\right]\;,\label{bieq}\\
\ddot{C}+k\dot{C}&=&e^{a\phi}\left[T_{\hat{c}\hat{c}}-\tau\rho\right]\;,\label{cieq}\\
\ddot{\phi}+k\dot{\phi}&=&\frac{1}{2}e^{a\phi}\left[T-(d-2)\tau\rho\right]\;,\label{phieq}\\
k^{2}&=&m\dot{B}^{2}+p\dot{C}^{2} + 2 e^{a\phi}\rho\;,\label{refk1} \\
k&\equiv& m\dot{B}+p\dot{C}-2\dot{\phi}\;.\label{refk2}
\eea
\end{subequations}

\noindent with $T$ being the trace of the total energy-momentum tensor defined as usual and $T_{\hat{b}\hat{b}}$ and $T_{\hat{c}\hat{c}}$ represent, in the orthonormal frame, the total pressure in the m-dimensional observed and p-dimensional compact spaces respectively.
The model considers two contributions to the total energy density 

\begin{subequations}\label{eq2}
\bea 
{\rm winding}\;\;\;\rho_{w}&=&\rho_{w}^{o} e^{-mB}\\
{\rm momentum}\;\;\;\rho_{m}&=&\rho_{m}^{o} e^{-mB-(1+p)C}
\eea
\end{subequations}

\no The equations admit the following solution for $m=3$ and $a=1$ (the dilaton coupling to D-branes from string theory),

\begin{subequations}
\bea 
C&=&C_{o}\;,\\
\phi&=&\phi_{o}\;,\\
B&=&\frac{2}{3}\ln(t)+B_{o}\;,
\eea
\end{subequations}

\noindent with

\begin{subequations}
\bea
e^{-(1+p)C_{o}}&=&\frac{\rho_{w}^{o}}{\rho_{m}^{o}}\left(\frac{p}{p+2}\right)\;,\\
e^{\phi_{o}-3B_{o}}&=&\frac{2}{3\rho_{w}^{o}}\left(\frac{p+2}{p+1}\right)\;.
\eea
\end{subequations}

We would like to point out the following points for our subsequent discussion

\begin{itemize}
\item{} The D-brane action in (\ref{eq1}) is not T nor S duality invariant (to be explicitly defined in the next chapter) with the matter terms given in (\ref{eq2}). So T and S duality invariances are explicitly broken by this model of D-branes.
\item{} A constant dilaton is admissible only for $m=3$ and this happens because for $m=3$ and $a=1$ the right hand side of the $\phi$ equation becomes proportional to the right hand side of the $C$ equation. Hence the stabilization of the extra dimensions makes the
right hand side of the $\phi$ equations identically zero. So it is the $C$ equation that triggers this. We will understand this effect in terms of S duality in the next section.
\item{} $\phi_{o}$ and $B_{o}$ are not separately fixed. Only the combination $\phi_{o}-3B_{o}$ is fixed by the constraint (zero-zero component of the tensor equation).
\end{itemize}

The solution is stable in the sense that if one perturbs the solution presented above, the variations can be found as follows

\begin{subequations}
\bea
\delta C&=&\frac{\alpha}{\sqrt{t}}\sin\left[\Omega\ln(t)+\beta\right]\;,\\
\delta B&=&-\frac{\gamma}{6}-\frac{\beta}{3t}\ln(t)\;,\\
\delta\phi&=&\frac{p\alpha}{2\sqrt{t}}\sin\left[\Omega\ln(t)+\beta\right]+\frac{\beta}{2t}-\frac{\gamma}{2}\;,
\eea
\end{subequations}

\noindent with $\alpha$, $\beta$ and $\gamma$ are small arbitrary constants and $\Omega=\sqrt{(4p+5)/3}/2$.

So no perturbation grows in time. Furthermore the constant arbitrary shifts on $B$ and $\phi$ are compatible with the initial constraint. That is,

\[
\delta\phi-3\delta B\to 0 \;\;{\rm as}\;\;t\to\infty\;,
\]

\no Which shows that the initial constraints for stabilization will not be altered. Therefor this model is fully compatible with a fixed dilaton. How this happens can be considered to be an accident but $m=3$ and $a=1$ are not arbitrary fine tunings of the parameters.

One would like to have a fixed dilaton for late time cosmologies in order to recover Einstein gravity. As recently argued \cite{newtr} with an explicit example S duality can actually help fix the dilaton (determine $\phi_{o}$) because it is a symmetry that directly acts on it. But in the mentioned work the stabilization of extra dimensions is achieved via the same element that stabilizes the dilaton. As this short review shows there
can be models which are compatible with a fixed dilaton and yet are solely responsible for the stabilization of extra dimensions. Even though Occam's razor would direct us to a more economical explanation we should nevertheless keep in mind that stabilization of extra dimensions and the dilaton can be achieved via different mechanisms. 

\section{General Formalism}

In what follows we will use a similar action 

\be{\label{eq:10}}
S=\int\;dx^{10}\sqrt{-g}\;e^{-2\phi}\left[R+4(\nabla\phi)^{2}+\mathcal{L}\right]\;.
\ee

\noindent But for the interaction  we will take the following generalization

\be{\label{eq:11}}
\mathcal{L}=-2\sum_{i}\;e^{a_{i}\phi}V_{i}(\phi)\;\rho_{i}\;,
\ee

\no with $\rho_{i}$ again having the fluid \footnote{It would appear we took $H_{\mu\nu\lambda}=0$. But we assume their effect can also be summarized as a hydrodynamical fluid} form. Also, as evident from our definition of the dilaton coupling to different $\rho$'s, $V_{i}(\phi)$'s should not be homogeneous in $e^{\phi}$. This will avoid any ambiguity in the definition of $a_{i}$'s.

The equations of motion following from (\ref{eq:10}) and (\ref{eq:11}) are

\begin{subequations}
\bea
\ddot{B}+k\dot{B}&=&\mathcal{B}-\mathcal{V'}\;,\\
\ddot{C}+k\dot{C}&=&\mathcal{C}-\mathcal{V'}\;,\\
\ddot{\phi}+k\dot{\phi}&=&\mathcal{F}-\frac{d-2}{2}\mathcal{V'}\;,\\
k^{2}&=&m\dot{B}^{2}+p\dot{C}^{2}+2\mathcal{E}\;,\\
k&\equiv&m\dot{B}+p\dot{C}-2\dot{\phi}\;.
\eea
\end{subequations}

\noindent with the following definitions

\begin{subequations}
\bea
\mathcal{E}&=&\sum_{i}e^{a_{i}\phi}\;V_{i}(\phi)\rho_{i}\;,\\
\mathcal{V'}&=&\frac{1}{2}\sum_{i}e^{a_{i}\phi}\frac{dV_{i}}{d\phi}\;\rho_{i}\;,\\
\mathcal{B}&=&\sum_{i}\;e^{a_{i}\phi}\;V_{i}\left[T^{i}_{\hat{b}\hat{b}}-\tau_{i}\rho_{i}\right]\;,\\
\mathcal{C}&=&\sum_{i}e^{a_{i}\phi}\;V_{i}\left[T^{i}_{\hat{c}\hat{c}}-\tau_{i}\rho_{i}\right]\;,\\
\mathcal{F}&=&\frac{1}{2}\sum_{i}e^{a_{i}\phi}\;V_{i}\left[T^{i}-(d-2)\tau_{i}\rho_{i}\right]\;.
\eea
\end{subequations}

\noindent We also have $\tau_{i}\equiv(a_{i}-2)/2$ and $d=1+m+p=10$.

Even though we implicitly started out with flat observed and internal dimensions the effect of curvature in these spaces can easily be accommodated within the formalism of the fluid approach. Quite generally we can simulate the effect of a curvature term via the following energy density

\be
\rho_{K}=-K\;q\;\exp\left[-(1+\omega_{K})mB-(1+\nu_{K})pC\right]\;,
\ee

\noindent where $q=\lbrace m,p\rbrace$ the dimensionality of the partition where there is curvature $K=\lbrace <0,0,>0 \rbrace$ representing hyperbolic, flat and spherical geometries respectively. The pressure coefficient of the partition having curvature will be $(2-q)/q$ and the other $-1$. For the curvature terms we of course have $a_{K}=0$. However one should not forget that this is just a sleight of hand. The curvature terms should be understood as such only at the equations of motion level not the action level. Since in reality they are part of $R$ (the marble) and not a part of some energy density in the interaction Lagrangian (the wood).

Furthermore a pure dilaton potential has $\omega_{p}=\nu_{p}=-1$ with an arbitrary potential $V_{p}(\phi)$ and $a_{p}$.

\section{Impact of T and S dualities}

\subsection{S duality}

The kinetic term in the action is invariant under S duality which is given by

\begin{subequations}
\bea
\phi&\to& -\phi\;,\\
g_{\mu\nu}&\to& e^{-\phi}g_{\mu\nu}\;.
\eea
\end{subequations}

\noindent In terms of our metric ansatz this means, 

\begin{subequations}
\bea
\phi&\to&-\phi\;,\\
B&\to& B-\phi/2\;,\\
C&\to& C-\phi/2\;.
\eea
\end{subequations}

The remaining part is not generally S invariant so we will have to impose conditions. As a remark we would like to point out that
S duality invariance cannot be imposed on the interactions as an interplay between two terms: this
would necessarily imply the equality of the respective pressure coefficients $\omega_{1}=\omega_{2}$ and $\nu_{1}=\nu_{2}$ meaning that the two terms can be combined with a new potential $V=V_{1}+V_{2}$. The effect of S duality invariance on each term gives the following condition

\begin{subequations}\label{sdual}
\bea
V_{i}(-\phi)&=&V_{i}(\phi)\;,\\
\tau_{i}&=&\frac{1}{8}\left[-1+m\omega_{i}+p\nu_{i}\right]\;.
\eea
\end{subequations}

\no This implies the following

\begin{itemize}
\item{} S duality invariance means $\mathcal{F}=0$. Thus we have a partial decoupling of the dilaton. This means that the conditions for $\phi$ stabilization can be orthogonal to the conditions for $C$ stabilization.
\end{itemize}
 
We now have a better understanding about why the model presented in the invitation is compatible with a fixed dilaton: it {\bf{\em{mimics}}} S duality invariance when $C$ is stabilized.

\subsubsection{Stabilization}

We can now look for general conditions for the stabilization of extra dimensions and the dilaton. We must have the following

\begin{subequations}
\bea
\mathcal{V'}(\phi_{o},C_{o}) &=& 0\;,\\
\mathcal{C}(\phi_{o},C_{o}) &=& 0\;.
\eea
\end{subequations}

A decoupling of the conditions will arise when $\mathcal{V'}(\phi_{o},C_{o}) = 0$ is achieved for

\be
\frac{dV_{i}}{d\phi}\vert_{\phi_{o}}=0\;.
\ee

\noindent We must point out here that in order not to also have a static $B$ forced on us by the conditions, one must either have all the $\omega_{i}$ equal to each other. Or, to at least have stabilization at late times, there must be two $\rho_{i}$ with the same $a_{i}$ and $V_{i}$ for the smallest $\omega$ in the model \footnote{The smallest $w$ terms will redshift slowest and will become the dominant terms at late times in cosmology.}. These two will in general be the momentum and winding modes of the object excited around extra dimensions. If these conditions are met $B$ will expand presumably according to a powerlaw solution depending on the smallest $\omega$ term.

To study the stability of this fixed point we can perturb the solutions and look how they evolve. The equation for the perturbations will be (we assumed a simple powerlaw for unperturbed solution for $B$)

\be
\delta\ddot{X}+\frac{\alpha}{t}\delta{\dot{X}}=-\frac{1}{t^{2}}\Sigma\delta X\;.
\ee

\noindent Here $\alpha$ is a model dependent number and $\delta X$ is a column vector with $\delta X^{T}=(\delta B,\delta C,\delta \phi)$. We call $\Sigma$ the stabilization matrix. It must have all positive eigenvalues for the fixed point to be stable. It has the following form for the model under study,

\begin{center}
% use packages: array
\[
\left(\begin{tabular}{lll}
$\Sigma_{BB}$ & $\Sigma_{BC}$ & $\Sigma_{B\phi}$ \\ 
0 & $\Sigma_{CC}$ & $\Sigma_{C\phi}$ \;.\\ 
0 & 0 & $\Sigma_{\phi\phi}$
\end{tabular}
\right)\]
\end{center}

The eigenvalues therefor are given by $\Sigma_{BB}$, $\Sigma_{CC}$ and $\Sigma_{\phi\phi}$ and they all have to be positive. The $\phi\phi$ part is relatively easy since the natural condition there just becomes

\be
\frac{d^{2}V_{i}}{d\phi^{2}}\vert_{\phi_{o}}>0\;.
\ee

The rest becomes a game related to whether stabilization of the extra dimensions can be achieved. However we would like to point out that the tridiagonal form of $\Sigma$ will prevent the mixing of $\delta C$ and $\delta B$ to $\delta\phi$ even if there is an instability in the evolution of the observed and extra dimensions. This comes as a special case circumventing the concerns presented in \cite{d1}. 

It is also rather nice to have dilaton stabilization via functions of the form $e^{a_{i}\phi}V_{i}(\phi)\rho_{i}$. The reason is that if $B$ is increasing (expanding observed universe) $\rho_{i}$ will decrease and it will act as a factor that damps the magnitude of the potentials. This damping effect in concert with ordinary friction term $k$ already present in the equations of motion (Hubble damping) will help reduce violent oscillations of the evolutions of the scale factors and the dilaton.

In \cite{newtr} a model which encompasses these general features have been presented. The protagonists for stabilization were $(m,n)$ strings with an action covariant in S duality. By this we mean that the action consisted of the winding and the momentum modes with the momentum parts manifestly S duality invariant, however the winding modes were not manifestly invariant by the transformation and an extra interchange of the quantum numbers $(m,n)\to(n,m)$ is required. This results in a spontaneous breaking of the S symmetry and $\phi_{o}$ was fixed to be $\ln(n/m)$.

\subsection{T duality}

T duality acts on the compact dimensions and the dilaton. For simplicity let us assume that we act it on all the p-dimensional compact space. This means we have

\begin{subequations}
\bea
C\to-C\;,\\
\phi\to\phi-pC\;.
\eea
\end{subequations}

This leaves the kinetic part invariant. A term in the interaction part will be T dual if

\begin{subequations}
\bea
a_{i}&=&2(1+\nu_{i})\\
V_{i}(\phi)&=&1
\eea
\end{subequations}

This implies, as expected, that such a term will not contribute to the righthand-side of the $C$ equations. 

Imposing further  S duality invariance condition in (\ref{sdual}) we get the TS self duality condition,

\be
\omega_{i}=\frac{1+(m-1)\nu_{i}}{m}\;.
\ee

There are various cases, for example $\nu_{i}=0$ meaning $\omega_{i}=1/m$ and $a_{i}=2$: some sort of fundamental string which behaves like radiation along observed dimensions and presureless dust along extra dimensions. For $\nu_{i}=-1$ we get $\omega_{i}=(2-m)/m$ and $a_{i}=0$. This case is analogous to a curvature term along the observed dimensions except that we must remember a {\em real} curvature term will have $\rho_{K}^{o}=-m K$. Thus such a term can have an interplay with a real curvature term.

\subsubsection{T duality invariance as an interplay between two terms}
One could argue that, unlike S duality, imposing T duality on a single term is not a must since its essence is about the contrast between winding and momentum modes. If one imposes T duality as an interplay between two terms one gets

\begin{subequations}
\bea
\omega_{1}&=&\omega_{2}\;,\\
a_{1}=a_{2}&=&2+(\nu_{1}+\nu_{2})\;,\\
V_{2}=V_{1}&=&1\;,\\
\rho_{1}^{o}&=&\rho_{2}^{o}\;.
\eea
\end{subequations}

\noindent An interesting case  is $\nu_{2}=-\nu_{1}$ giving $a_{i}=2$: winding and momentum modes of a fundamental string. This results in the following equations

\begin{subequations}
\bea
\ddot{B}+k\dot{B}&=& 2\omega f(\phi,B,C)\cosh(\nu pC)\;,\\
\ddot{C}+k\dot{C}&=& -2\nu f(\phi,B,C)\sinh(\nu pC)\;,\\
\ddot{\phi}+k\dot{\phi}&=& (m\omega-1)f(\phi,B,C)\cosh(\nu pC)+\nonumber\\
&&-p\nu f(\phi,B,C)\sinh(\nu pC)\;,
\eea
\end{subequations}

\noindent with $f(\phi,B,C)=\rho^{o}\;e^{2\phi-(1+\omega)mB-pC}$. As evident the radion will stabilize at $C=0$ and this gets rid of one term in the $\phi$ equations. The system will be compatible with a fixed dilaton if $\omega=1/m$ which is
like radiation along observed dimensions. This example is like the one presented in the invitation: $C$ stabilization {\em{\bf mimics}} S duality invariance for a specific albeit physically meaningful choice of parameters.

\section{A note on the cosmological constant and Acceleration}

A pure cosmological constant has $a_{\Lambda}=0$ and $\omega_{\Lambda}=\nu_{\Lambda}=-1$ \footnote{The reason for choosing $a_{\Lambda}=0$ for a cosmological constant is that $\Lambda$ must come as an addition to $R$ in the action}. Therefor this term is T invariant but  not S invariant. The equations read

\begin{subequations}
\bea
\ddot{B}+k\dot{B}&=& 0 \;,\\
\ddot{C}+k\dot{C}&=& 0 \;,\\
\ddot{\phi}+k\dot{\phi}&=& -\Lambda\\
k^{2}&=&m\dot{B}^{2}+p\dot{C}^{2}+2\Lambda\;.
\eea
\end{subequations}

The solutions are $B=B_{o}$, $C=C_{o}$ and $\phi=\pm\sqrt{\Lambda/2}\;t$ \footnote{In string theory $\Lambda$ (or $c$ in \cite{a1}) can be negative. In that case the solutions are different.}. Therefor the string coupling $g_{s}=e^{\phi}$ is accelerating and the spatial dimensions are static. However this is the picture in the string frame. When one converts to Einstein frame one recovers  only $B_{E}=\ln(t_{E})+B_{oE}$ and $C_{E}=\ln(t_{E})+C_{oE}$. These are not accelerating. This curious result shows that it is not always possible to get acceleration with a cosmological constant.

One can think of an S invariant pure dilaton potential term without a constant accompanying $V$. This will have $\tau_{\Lambda_{\phi}}=-5/4$, $a_{\Lambda_{\phi}}=-1/2$ along with $\omega_{\Lambda_{\phi}}=\nu_{\Lambda_{\phi}}=-1$ and $V_{\Lambda_{\phi}}=\Lambda_{\phi}$. The equations read in this case

\begin{subequations}
\bea
\ddot{B}+k\dot{B}&=& \frac{1}{4}\Lambda_{\phi}\;e^{-\phi/2}\;,\\
\ddot{C}+k\dot{C}&=& \frac{1}{4}\Lambda_{\phi}\;e^{-\phi/2}\;,\\
\ddot{\phi}+k\dot{\phi}&=& 0\;,\\
k^{2}&=&m\dot{B}^{2}+p\dot{C}^{2}+2\Lambda_{\phi}\;e^{-\phi/2}\;. 
\eea
\end{subequations}

\noindent These are compatible with a constant dilaton and hence the Einstein frame and the string frame becomes identical. Therefor $\Lambda_{\phi}e^{-\phi_{o}/2}$ acts like an effective cosmological constant in Einstein gravity and we recover accelerating solutions. 
It is interesting to speculate on an interplay between $\Lambda_{\phi}$ and $\Lambda$ the former being only S invariant and the latter being only T invariant. In these cases one recovers the following equations

\begin{subequations}
\bea
\ddot{B}+k\dot{B}=&&\frac{1}{4}\Lambda_{\phi}\;e^{-\phi/2}\;,\\
\ddot{C}+k\dot{C}=&& \frac{1}{4}\Lambda_{\phi}\;e^{-\phi/2}\;,\\
\ddot{\phi}+k\dot{\phi}=&& -\Lambda\;,\\
k^{2}=&&m\dot{B}^{2}+p\dot{C}^{2}+2\Lambda_{\phi}\;e^{-\phi/2}+2\Lambda \;.
\eea
\end{subequations}

Of course in this scenario one cannot stabilize the dilaton. However if initially we are in a regime where the dilaton is decreasing as $\phi=-\sqrt{\Lambda/2}t$ (similar to the first example in this section) the scale factors will evolve as if they are under the influence of an ever increasing cosmological constant. This is an interesting example in that it could require much less time to inflate to the required number of e-foldings. A saner scenario will be  a non-constant dilaton S invariant potential $V(\phi)$ along with a cosmological constant.

\begin{subequations}
\bea
\ddot{B}+k\dot{B}=&&\frac{1}{4}\;e^{-\phi/2}V-\frac{1}{2}\;e^{-\phi/2}V' \;,\\
\ddot{C}+k\dot{C}=&& \frac{1}{4}\;e^{-\phi/2}V-\frac{1}{2}\;e^{-\phi/2}V' \;,\\
\ddot{\phi}+k\dot{\phi}=&& -\Lambda -2e^{-\phi/2}V'\;,\\
k^{2}=&&m\dot{B}^{2}+p\dot{C}^{2}+2e^{-\phi/2}V+2\Lambda\;. 
\eea
\end{subequations}

The dilaton acts as if it is under the influence of an effective potential 

\be
\frac{dV_{eff}}{d\phi}=\Lambda+2e^{-\phi/2}\frac{dV}{d\phi}\;.
\ee

It is apparent that since $\Lambda$ explicitly breaks S invariance it is not possible to stabilize the dilaton at $\phi_{o}=0$  even if $V$ does not spontaneously break S symmetry. If the dilaton can be stabilized the scale factors will evolve as if they are under the influence of an effective cosmological constant $\Lambda_{eff}=\Lambda+e^{-\phi_{o}/2}V(\phi_{o})$. But we emphasize again that this is due to the presence of $V_{\phi}$, if it is absent we will be back to the case at the beginning of this chapter and one will not have acceleration.

\section{Conclusion}

The key observation we have presented is that {\bf T-duality invariant actions can for certain physically relevant cases mimic S duality invariance when internal dimensions stabilize} and thus the resulting model will be equivalent to Einstein gravity since one has a constant dilaton (unspecified value) and stabilized radion.

Another important result that follows from our discussions and that justifies and strengthens a known fact is that it is in general wrong to assume that the dilaton can be taken to
be a constant a priori. We have shown various examples of this and possibly the most striking one is the impossibility of a constant dilaton for a cosmological constant. The dilaton can be stabilized with a potential and if the potential has a non-zero value at its minimum one recovers the ordinary interpretation of vacuum energy as cosmological constant.

We have made use of the most general possible interaction terms which are in the form of hydrodynamical fluid. We have seen that S duality invariance has to be imposed term by term in the interaction. Consequently the dilaton equation is partially decoupled from the other equations and this is great help in studying radion and dilaton stabilization. T duality invariance on the other hand can be imposed term by term or as an interplay between two terms in the Lagrangian. If a single term in the interaction is T duality invariant it will not
contribute to the evolution of the extra dimensions and hence is compatible with a constant radion albeit of unspecified value. If the sum of two terms is invariant under the T symmetry radion will evolve and will be stabilized at $C=0$. We have shown some physically interesting cases of this sort which also triggers {\em mimicking} of S duality and the stabilization of the radion field results in a theory compatible with a constant dilaton and hence one recovers Einstein's theory. It is also possible to have the opposite form: an S invariant action stabilizing the dilaton -that is fix the value at $\phi_{o}=0$ (or $\phi_{o}\neq0$ if there is room for spontaneous breaking of S symmetry)- and being compatible with a constant albeit unspecified value of the radion. 

To conclude it is of crucial importance to know what mechanism is actually responsible for the stabilization of the dilaton and also whether this mechanism also triggers the stabilization of the dilaton or vice-versa. We see that T, S and TS invariant interactions provide a good startin point to resolve this issue

\section{Acknowledgements}
The author thanks A. Kaya and M. Ar{\i}k for various fruitful discussions on the subject.

\end{document}